\documentclass[a4paper,11pt]{article}
\usepackage{pos}

\newcommand{\dHybridR}{{\tt dHybridR}}

\newcommand{\vb}{v_{\rm bst}}
\newcommand{\gb}{\gamma_{\rm bst}}

\title{The saturation of the Bell instability and its implications for cosmic ray acceleration and transport}
 \ShortTitle{Bell's Instability Saturation}

\author*[a,b]{Damiano Caprioli}
\author[a]{Georgios Zacharegkas}
\author[b]{Colby C.~Haggerty}
\author[a]{Siddhartha Gupta}
\author[a]{Benedikt Schroer}

\affiliation[a]{Department of Astronomy and Astrophysics, University of Chicago, 5640 S Ellis Ave, Chicago, IL 60637, USA}
\affiliation[b]{Enrico Fermi Institute, University of Chicago, 5640 S Ellis Ave, Chicago, IL 60637, USA}
\affiliation[c]{Institute for Astronomy, University of Hawaii, Honolulu, HI 96822, USA}

\emailAdd{caprioli@uchicago.edu}
\abstract{
The non-resonant (Bell) streaming instability driven by energetic particles is crucial for producing amplified magnetic fields that are key to the acceleration of cosmic rays (CRs) in supernova remnants, around Galactic and extra-galactic CR sources, and for the CR transport. We present a covariant theory for the saturation of the Bell instability, substantiated by self-consistent kinetic simulations, that can be applied to arbitrary CR distributions and discuss its implications in several heliospheric and astrophysical contexts.
}

\ConferenceLogo{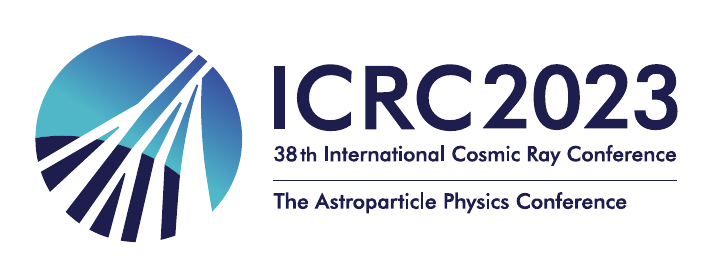}

\FullConference{%
38th International Cosmic Ray Conference (ICRC2023)\\
  26 July - 3 August, 2023\\
  Nagoya, Japan}

\begin{document}
\maketitle

\section{Introduction}
Collisionless shocks associated with supernova remnants (SNRs) are believed to be the primary source of Galactic cosmic rays (CRs) up to the ``knee", at rigidities of a few PV.
Particles are energized via diffusive shock acceleration (DSA) \cite{bell78a,blandford+78}, which requires CRs to be confined close to the shock, and hence strong, turbulent magnetic fields \cite{blasi+07}. 
Magnetic turbulence and acceleration are closely related and the study of the growth and saturation of CR-driven instabilities is crucial to explain the origin of high-energy CRs. 

While the left-handed modes driven by the resonant CR streaming instability \citep{kulsrud+69,zweibel79, achterberg83} were initially thought to be the only ones important for the self-confinement of CRs, Bell \cite{bell04} showed that the right-handed modes may grow faster and saturate at much larger amplitudes.
Such a \emph{nonresonant streaming instability}, (or ``Bell instability"), is crucial for the dynamics of non-relativistic shocks \citep[e.g.,][]{haggerty+20,caprioli+20}, as well as for the determining the highest energy achievable in SNRs \citep[e.g.,][]{caprioli+13,caprioli+14a,caprioli+14b,caprioli+14c,crumley+19}.

The linear theory shows that the Bell instability grows faster than the resonant instability when the maximally unstable wavelength is much smaller that the CR gyroradius, i.e., $k_{\rm max} r_L\gg 1$. 
This happens for large CR currents/energy densities (as discussed, e.g., in \S4.3 of \cite{bell04} and \S3 of \cite{amato+09}),  provided that the growth rate does not exceed $\Omega_{ci}$, the gyrofrequency of thermal ions. 
In this regime right-handed, circularly-polarized modes are driven unstable and the wavenumber and the growth rate of the fastest-growing mode read
\begin{equation}\label{eqth:k_max Bell}
	k_{\max} = \frac{4\pi}{c} \frac{J_{\rm cr}}{B_0} = \frac{1}{2} \left( \frac{n_{\rm cr}}{n_{\rm g}} \right) \left( \frac{v_{\rm d}}{v_{\rm A,0}} \right) d_i^{-1};\quad
	\gamma_{\max} = k_{\max} v_{\rm A,0} = \frac{1}{2} \left( \frac{n_{\rm cr}}{n_{\rm g}} \right) \left( \frac{v_{\rm d}}{v_{\rm A,0}} \right)\Omega_{ci} \; ,
\end{equation}
where $e$ and $m$ are the proton charge and mass, $J_{\rm cr}=e n_{\rm cr}v_{\rm d}$ is the CR current, $n_{\rm cr}$ and $v_{\rm d}$ are the CR number density and drift velocity relative to the background plasma, $B_0$ and $n_{\rm g}$ are the background magnetic field and plasma number density, $v_{\rm A,0} \equiv B_0/\sqrt{4\pi m n_{\rm g}}$ is the initial Alfv\'{e}n speed, $\Omega_{ci} \equiv eB_0/(mc)$, and $d_i \equiv v_{\rm A,0}/\Omega_{ci}$ is the ion inertial length. 

Note that, depending on the CR distribution, many differnt modes, both parallel and transverse to the magnetic field may grow, as discussed in the thorough review by Bret \cite{bret09}, where the growth rate of Weibel, two-stream, Buneman, filamentation, Bell, and cyclotron instabilities are compared.
Nevertheless, for most astrophysical applications, and surely for SNR shocks, the Bell instability is the most prominent channel for generating magnetic turbulence \citep[e.g.,][]{caprioli+14a,caprioli+14b, caprioli+18, haggerty+19a, orusa+23}.

%The original derivation of the growth rate of the Bell instability did not account for situations where the background plasma is not strictly cold or collisionless.
%The Bell instability is expected to be modified when the background plasma is sufficiently warm; 
%in this regime (dubbed ``WICE'' for warm ions/cold electrons) the fastest growing wavenumber shifts to $k_{\rm wice}<k_{\max}$ and the growth rate is suppressed \citep{reville+08a,zweibel+10,marret+21}.
%Additionally, the instability is modified in systems where the collisional time scales becomes comparable to the growth rate \citep{reville+07}. 
%Systems with proton-neutral collisions were found to have reduced growth rates and saturated magnetic field amplitudes, while systems with proton-proton collisions resulted in a larger saturated magnetic field, owing to the suppression of temperature anisotropy generation \citep{marret+22}.
%While these consideration are potentially important to select systems, they are not included here.

\subsection{Simulating the Bell Instability}
MHD simulations have shown that, for a fixed CR current, large amplification factors can be achieved \citep[e.g.,][]{bell04,bell05, matthews+17}.
Nevertheless, these simulations cannot self-consistently capture the backreaction of the growing modes on the CRs, hence they cannot be used to assess the saturation of the Bell instability.
Particle-in-cell (PIC) simulations \citep{riquelme+09, ohira+09b, gargate+10, gupta+21} confirmed that for typical CR currents the Bell instability grows as expected and saturates to levels of $\delta B/B_0 \gg1$, unless the current is too strong \citep{niemiec+08}, in which case a transverse filamentary mode grows faster than Bell.
Such a saturation was ascribed to the background plasma being accelerated in the direction of the CR drift velocity, which reduces the effective CR current $\mathbf{J}_{\rm cr}$ \cite{riquelme+09,gargate+10}.

Reville et al.~\cite{reville+13} used a MHD+Vlasov code to run \emph{driven simulations} in which the CR current is kept constant, a situation more akin to a shock precursor, where the upstream plasma is constantly exposed to ``fresh'' CRs; 
in this work the growth of the field was time-limited, and no saturation of the magnetic field was achieved.
Kobzar et al.~\cite{kobzar+17} performed PIC 2D simulations in a non-periodic box to follow the spatio-temporal evolution of the instability, though the very large current they used led to the formation of a shock, which should not happen for arbitrary CR distributions.

In our work \cite{zacharegkas+22}, we study the Bell instability via hybrid simulations using the massively parallel code \dHybridR~\citep{gargate+07,haggerty+19a}, where ions are treated kinetically and their relativistic dynamics is retained. 
We perform {\it driven simulations} in which CRs are injected in the simulation box at a constant rate on the left side and are free to leave from the right, while the thermal background plasma and the electro-magnetic fields are subject to periodic boundary conditions.
This setup allows for a self-consistent coupling between CRs and thermal plasma, which eventually leads to the saturation of the instability. 
We explore a large range of parameters that characterize the CR current, always in the regime in which Bell is the fastest growing instability, and use a suite of 1D, 2D, and 3D simulations to investigate how the amplified magnetic field at saturation scales with the CR parameters.
%We provide the first simulation-validated prescription for the level of the final magnetic field at saturation.
We remand to the full paper for the details of the runs and the discussion of the linear and non-linear stages of our benchmark runs, as well as for convergence studies in number of spatial dimensions, box size, particles per cells, and space/time resolution. 

In these proceedings we summarize our main finding, namely the formula that gives the amplified magnetic field at saturation as a function of the initial CR net momentum flux and compare such an expression with the original ansatz put forward by Bell based on a heuristic argument. 
We briefly discuss the implications of this new saturation value, which is typically smaller that Bell's, for astrophysical applications, viz., the maximum energy achievable in SNRs.

\section{A General Formula for the Saturation of the Bell Instability}

\begin{figure}[t]
\centering
\vspace{-5mm}
\includegraphics[width=0.7\columnwidth]{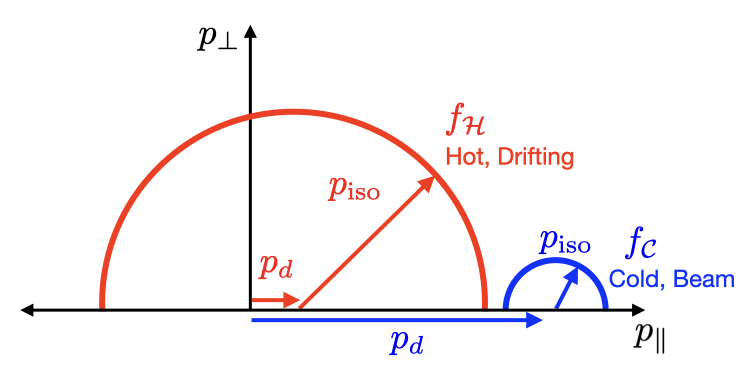} 
\vspace{-5mm}
\caption{Diagram of the initial CR distributions, distinguishing the hot and cold cases.}
\label{fig:distro diag}
\end{figure}

We consider CRs with number density $n_{\rm cr}$ and isotropic monochromatic momentum $p_{\rm iso} \equiv \gamma_{\rm iso} m v_{\rm iso}$ in their rest frame, which drift with velocity $\mathbf{v}_{\rm d}=v_{\rm d}\hat{\mathbf{e}}_x$ relative to the thermal plasma;
this corresponds to a current $\mathbf{J}_{\rm cr} = e n_{\rm cr} \mathbf{v}_{\rm d}$; also, be  $\mathbf{B} = B_0\hat{\mathbf{e}}_x$ the initial magnetic field.

While the CR current, and hence the growth rate, only depend on $n_{\rm cr}$ and $v_{\rm d}$, the CR net momentum and energy fluxes depend on $p_{\rm iso}$, too.
Therefore, we may have two limiting cases: one in which the CRs are a cold beam with $p_{\rm d}\gg p_{\rm iso}$, or a hot drifting distribution with $p_{\rm d}\ll p_{\rm iso}$ (Fig.~\ref{fig:distro diag}).
The goal of this work is to provide a prescription for the amplified magnetic field related to the free energy/momentum flux in the initial CR distribution.

If we look at the saturation prescription suggested by Bell \cite{bell04} \citep[see also][]{blasi+15}, in the limit in which CRs are relativistic and the drift is not, the final amplification reads:
\begin{equation}\label{eq:xibell}
    \xi_{\rm Bell}\equiv \left(\frac{\delta B}{B_0}\right)^2 =  \frac{U_{\rm cr}}{U_{\rm B}}\frac{v_{\rm d}}{2c} = \eta \gamma_{\rm iso}\frac{ v_{\rm d}c}{v_A^2};
\end{equation} 
where $U_{\rm cr}\equiv\gamma_{\rm iso}m n_{\rm cr}c^2$ and $U_{\rm B}\equiv m n_{\rm g}v_A^2/2$ are the CR and magnetic energy densities and we introduced $\eta\equiv n_{\rm cr}/n_{\rm g}$.
Also note that, if one poses $p_\perp\approx \gamma_{\rm iso}mc$ (hot, relativistic CRs), then
\begin{equation}\label{eq:kmaxrl}
     k_{\rm max} r_{\rm L} = \eta\frac{v_{\rm d}}{v_A}\frac{p_\perp}{mv_A} \simeq \xi_{\rm Bell}.
\end{equation}
The saturated magnetic field is thus related to the value of $k_{\rm max} r_{\rm L}\gg 1$, in the sense that Bell's \emph{ansatz} is equivalent to asking that at saturation  $k_{\rm max}(\delta B) r_{\rm L}(\delta B)\sim 1$.

%For a cold beam, instead, one may argue that it would be more reasonable to pose $p_\perp \sim p_{\rm d}$ in Equation \ref{eq:kmaxrl}, since when modes go non-linear the CR gyroradii should not care about the direction of the magnetic field, which returns 
%\begin{equation}\label{eq:xibellcold}
%    \xi_{\rm Bell}^{\rm (cold)} = \eta \gamma_d\frac{v_{\rm d}^2}{v_A^2}.
%\end{equation}

Eq.~\ref{eq:xibell} is deceitfully similar to the ratio of CR and magnetic \emph{energy fluxes}, but the denominator is \emph{not} the magnetic energy flux (waves do not move at $c$);
this raises the questions of what are the physical quantities that balance out at saturation, and what is the covariant expression that encompasses and generalizes Eq.~\ref{eq:xibell} for arbitrary CR distributions.

\begin{figure}
    \centering
    \vspace{-5mm}
    \includegraphics[width=0.8\textwidth]{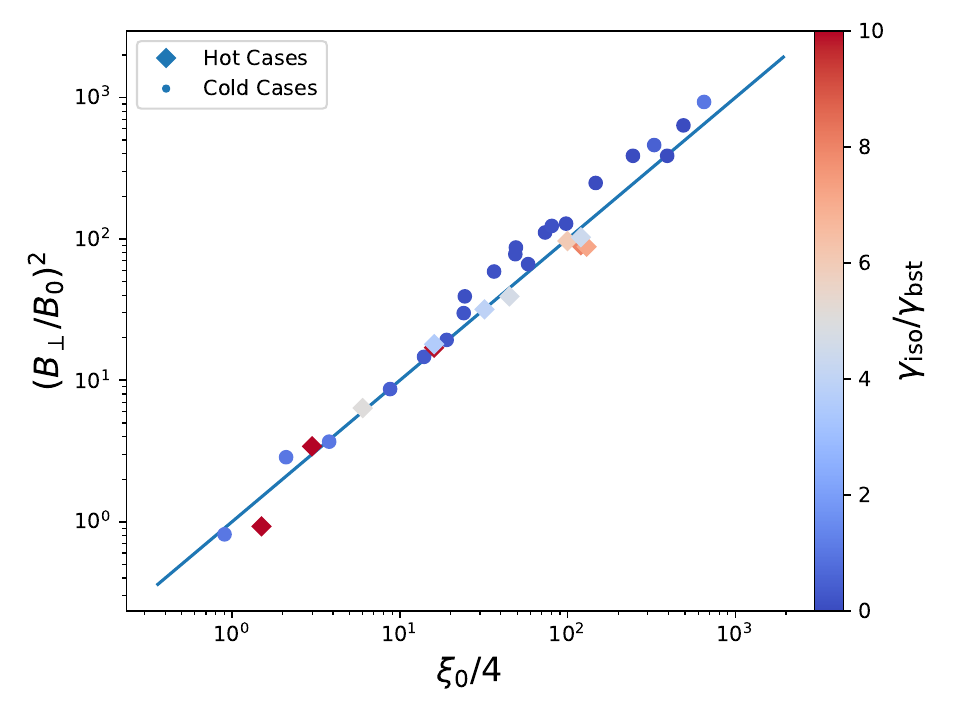}
    \vspace{-5mm}
    \caption{Scaling of the self-generated magnetic field with the parameter introduced in Eq.~\ref{eq:xi}, $\delta B^2/B_0^2\simeq \xi/4$, for many different hot/cold CR currents (see \cite{zacharegkas+22} for more details about the runs).}
    \label{fig:cases}
\end{figure}

In the CR rest frame, the CR mass density is $\tilde{\rho}_{\rm cr}$ and their total (including the rest mass) energy density and isotropic pressure are $e_{\rm cr}= \gamma_{\rm iso} \tilde{\rho}_{\rm cr}c^2$ and $p_{\rm cr}=\frac13 \gamma_{\rm iso}\tilde{\rho}_{\rm cr}c^2\beta_{\rm iso}^2$.
In an arbitrary frame that moves with velocity $\mathbf{v}_{\rm bst}$ and has a corresponding Lorentz factor $\gb$, the CR density is $\rho_{\rm cr}=\gb\tilde{\rho}_{\rm cr}$ and the components of the CR stress tensor read \citep[see, e.g., \S133  of][]{landau6}:
\begin{equation}
T^{\mu\nu}= (e_{\rm cr}+p_{\rm cr}) u_{\rm bst}^\mu u_{\rm bst}^\nu + p_{\rm cr}\eta^{\mu\nu},
\end{equation}
where $u_{\rm bst}^\mu$ is the four-velocity constructed with $\mathbf{v}_{\rm bst}$  and $\eta^{\mu\nu}$ is the Minkovski metric. 
Note that, if CRs are relativistic, then $\vb$ is generally \emph{larger} than the CR drift velocity in the final frame; 
the two are connected (see, e.g., \cite{gupta+21}) and since simulations are setup with an effective boost, we will give provide the saturation as a function of $\vb$ rather than $v_{\rm d}$.

%$$T^{00} = \Gamma^2 (e+p) - p \; ,$$
%$$T^{0i} = T^{i0}=\Gamma^2 (e+p) \frac{v_b^i}{c} \; ,$$
%$$T^{ij} = \Gamma^2 (e+p) \frac{v_b^iv_b^j}{c^2}+p\delta^{ij}$$where $i,j=1,2,3$ and $\delta^{ij}$ is the Kronecker delta symbol. 
%Here $T^{00}$ has the usual meaning of energy density, while $T^{ij}$ is the density of flux of momentum $p_j$ along the $i$ direction;
%$T^{0i}$ does not have a non-relativistic counterpart, though $T^{0i}/c$ and $cT^{0i}$ are the density of momentum along $i $ and the energy flux along $i$, respectively. 

%The questions that arises is: which component(s) of $T$ is the saturation of the Bell instability related to? 

We remand to the full paper \cite{zacharegkas+22} for the explanation of how we figured out the components of $T^{\mu\nu}$ that matter and  report here just the final result.
Fig.~\ref{fig:cases} shows that the transverse, self-generated, component of the magnetic field at saturation nicely correlates with the following quantity:
\begin{equation}\label{eq:xi}
    \xi= \frac{T^{11}-p_{\rm cr}}{P_{B,0}} = 
    \frac{T^{01}}{P_{B,0}} \frac{v_{\rm bst}}{c} =
    \frac{T^{00}-\tilde{\rho}_{\rm cr}c^2}{P_{B,0}} =
     2 \gamma_{\rm iso}\gamma_{\rm bst} \frac{n_{\rm cr}}{n_{\rm g}} \frac{v_{\rm bst}^2}{v_A^2} \left( 1+\frac{1}{3}\frac{v_{\rm iso}^2}{c^2}\right),
\end{equation}
with the final amplified field being $\delta B^2/B_0^2\simeq \xi_0/4$.
Here $\xi_0$ has the meaning of the net (i.e., anisotropic) momentum flux in the CRs (see the terms with $T^{11}$ and $T^{01}$), or equivalently of the density in kinetic energy in the CR drift (term with $T^{00}$), normalized to the initial magnetic pressure.
The prescription is validated by 1D, 2D, and 3D simulations, listed in \cite{zacharegkas+22}, with many different CR distributions, both in the hot and cold regimes (color code and legend in Fig.~\ref{fig:cases}).

%\subsection{Comparison with Bell's Ansatz}

\begin{figure}
    \centering
    \vspace{-5mm}
    \includegraphics[width=0.7\textwidth]{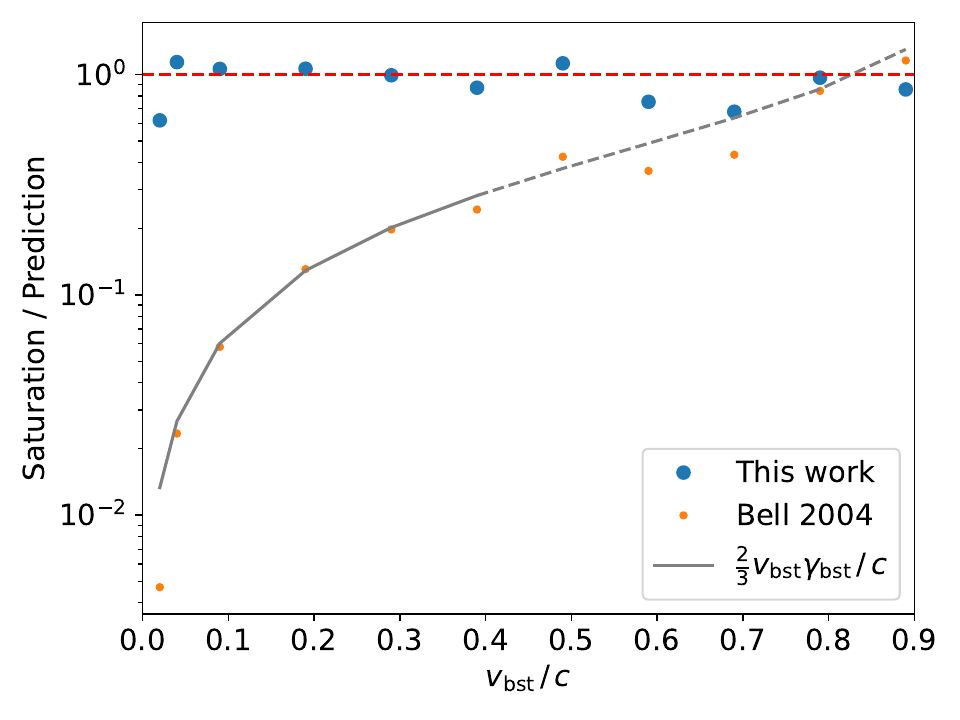}
    \vspace{-5mm}
    \caption{Comparison of Bell's original prescription for saturation with the one in Eq.~\ref{eq:xi}; the dashed line indicates the extrapolation of the non-relativistic boost regime.
    Our simulations suggest that the system saturates when it runs out of CR momentum flux, rather than energy flux.}
    \label{fig:ansatz}
\end{figure}

In the Bell limit (relativistically hot, non-rel drift), Eq.~\ref{eq:xi} reduces to:
\begin{equation}
    \xi_0\simeq  \gamma_{\rm iso}\eta \frac{v_{\rm bst}^2} {v_A^2}=
    \xi_{\rm Bell} \frac{2}{3}\gamma_{\rm bst}\frac{v_{\rm bst}}{c},
\end{equation}
which (except for a factor of $\sim 2$) corresponds to a saturation smaller than Bell's ansatz by a factor  $v_{\rm bst}/c$.
The physical meaning of this result is that it is more likely for the instability to quench because of the system running out of net momentum (pressure), rather than of energy flux. 
Fig.~\ref{fig:ansatz} confirms that this is the case: the new prescription for $\xi$ (Eq.~\ref{eq:xi}) provides a better agreement with the saturated $B_\perp$ than Eq.~\ref{eq:xibell}, the discrepancy being largest for smaller values of $v_{\rm bst}/c$.

At first sight, this result may exacerbate the issue of accelerating CRs up to the knee in SNRs, which already with Bell's prescriptions comes short of about one order of magnitude for typical remnants \cite[e.g.,][]{bell+13,cardillo+15,cristofari+22}.
However, one has to remember that magnetic field amplification in SNRs occurs both in the precursor, due to diffusing CRs (a hot distribution with $v_{\rm d}\sim v_{\rm sh}\ll c$), and far upstream, due to escaping CRs (a rather cold beam with $v_{\rm d}\sim c$) \cite[e.g.,][]{caprioli+09a,caprioli+14b}.
Escaping CRs are fewer in number, by a factor of $\sim v_{\rm sh}/c$ for a $p^{-4}$ distribution, but maximize the drift velocity, having $v_{\rm d}\sim c$; 
since $\xi_0\propto v_{\rm d}^2$, the net result is that escaping CRs are predicted to amplify the field more than diffusing ones, at a level comparable with Bell's ansatz when considering diffusing CRs.

\section{Conclusions}\label{sec:conclusions}
We have used controlled hybrid simulations to investigate the saturation of the Bell instability \citep{bell04} for a wide range of CR distributions, spanning from cold beams to hot-drifting cases (Fig.~\ref{fig:distro diag}).
We used a suite of 1D, 2D, and 3D simulations to assess the final amplitude of the self-generated magnetic field and our main finding is that what controls the saturation is the net momentum flux in CRs, not their energy flux (Eq.~\ref{eq:xi}).
This suggests that in shocks most of the amplification must be driven by escaping particles, rather than by CRs diffusing in the precursor, but it does not change the expectation for the highest energy CRs may achieve in SNRs.   

\vspace{2mm}
Simulations were performed on resources provided by the University of Chicago Research Computing Center and on  TACC's Stampede2 via the ACCESS allocation TG-AST180008.
We wholeheartedly thank Ellen Zweibel, Pasquale Blasi and Elena Amato for interesting and stimulating discussions.
D.C. was partially supported by NASA through grants 80NSSC20K1273 and 80NSSC18K1218 and NSF through grants AST-1909778, PHY-2010240, and AST-2009326, while C.C.H was supported by NSF FDSS grant AGS-1936393 and NASA grant 80NSSC20K1273.

\bibliographystyle{unsrt}
\bibliography{Total.bib}

\end{document}